\documentclass[reprint,notitlepage,amsmath,amsfonts,amssymb,floatfix,superscriptaddress]{revtex4-2}
\usepackage[export]{adjustbox}
\usepackage{bm}
\usepackage{color}
\usepackage{tikz-cd}
\usepackage{braket}
\usepackage[normalem]{ulem}

\newcommand{\T}{\mathsf{T}}

\begin{document}
\title{Hartree-Fock-Bogoliubov theory for number-parity--violating fermionic Hamiltonians}
\author{Thomas M. Henderson}
\affiliation{Department of Chemistry, Rice University, Houston, TX 77005-1892}
\affiliation{Department of Physics and Astronomy, Rice University, Houston, TX 77005-1892}

\author{Shadan Ghassemi Tabrizi}
\affiliation{Department of Chemistry, Rice University, Houston, TX 77005-1892}

\author{Guo P. Chen}
\affiliation{Department of Chemistry, Rice University, Houston, TX 77005-1892}

\author{Gustavo E. Scuseria}
\affiliation{Department of Chemistry, Rice University, Houston, TX 77005-1892}
\affiliation{Department of Physics and Astronomy, Rice University, Houston, TX 77005-1892}
\date{\today}

\begin{abstract}
 It is usually asserted that physical Hamiltonians for fermions must contain an even number of fermion operators. This is indeed true in electronic structure theory.  However, when the Jordan-Wigner transformation is used to map physical spin Hamiltonians to Hamiltonians of spinless fermions, terms which contain an odd number of fermion operators may appear.  The resulting fermionic Hamiltonian thus does not have number parity symmetry, and requires wave functions which do not have this symmetry either.  In this work, we discuss the extension of standard Hartree-Fock-Bogoliubov (HFB) theory to the number-parity--nonconserving case.  These ideas had appeared in the literature before, but, perhaps for lack of practical applications, had to the best of our knowledge never been employed.  We here present a useful application for this more general HFB theory based on coherent states of the SO(2$M$ + 1) Lie group, where $M$ is the number of orbitals.  We also show how using these unusual mean-field states can provide significant improvements when studying the Jordan-Wigner transformation of chemically relevant spin Hamiltonians.
\end{abstract}

\maketitle

\section{Introduction}
Hamiltonians may have symmetries which can be used to label states.  At the mean-field level, these symmetries can be spontaneously broken, which signals the breakdown of the symmetry-adapted mean-field picture and the onset of strong correlation.  We can take advantage of this phenomeon to build a fundamentally mean-field picture of some strongly-correlated problems through the use of symmetry-projected mean-field methods.\cite{Lowdin1955c,Ring1980,Blaizot1985,Schmid2004,JimenezHoyos2011b,JimenezHoyos2012,Sheikh2021}

Many of these symmetries are familiar to us: number symmetry, spin symmetry, and point group symmetry, for example.  Occasionally we make use of more esoteric molecular symmetries such as complex conjugation or time reversal.  In this work, we discuss a symmetry which physical fermionic Hamiltonians generally possess but which we almost never consider.  That symmetry is number parity
\begin{equation}
\Pi = \mathrm{e}^{\mathrm{i} \, \pi \, N},
\end{equation}
where $N$ is the fermionic total number operator, and it encompasses the notion that physical fermionic Hamiltonians are written in terms of an even total number of creation and annihilation operators.  So ubiquitous is this symmetry that we take it for granted, and rarely discuss it at all.  Number eigenstates are also number parity eigenstates (but not necessarily \textit{vice versa}); systems with even particle number are eigenstates of $\Pi$ with eigenvalue $+1$ and those with odd particle number are eigentates with eigenvalue $-1$.  More generally, $\Pi$ eigenstates with eigenvalue $+1$ are linear combinations of states with even particle number and those with eigenvalue $-1$ are linear combinations of states with odd particle number.  Usually in quantum chemistry we work with states with good particle number, in which case electronic excitation preserves number parity while ionization or electron attachment processes change it.

We had long dismissed number parity symmetry breaking as essentially irrelevant.  After all, in our work number parity symmetry never spontaneously breaks at the mean-field level because our mean-field wave functions explicitly cannot break it.

Recently, however, we began to consider using the Jordan-Wigner transformation\cite{Jordan1928} or extensions of it\cite{Wang1990,Henderson2023} to transform physical spin Hamiltonians into fermionic ones which we then solve with fermionic mean-field methods.\cite{Henderson2022,Chen2023,Henderson2023}  Particularly when these Hamiltonians have spin frustrations, we may construct fermionic Jordan-Wigner--transformed Hamiltonians which do not possess number parity symmetry.  This is because when we write even an $S_z$-conserving spin Hamiltonian in a noncollinear basis, terms like $S_p^z \, S_q^+$ appear, and upon JW-transformation they map into operators like $n_p \, a_q^\dagger$ (vide infra) which do not commute with $\Pi$.  We would therefore have to solve these Hamiltonians with number parity violating mean-field methods.

This problem has been tackled before.  In 1977,\cite{Fukutome1977,Fukutome1977b} Fukutome and coworkers proposed a generalization of Hartree-Fock-Bogoliubov (HFB) theory to the number parity violating case.  Their method is based on a standard HFB linear canonical transformation for $M$ orbitals, which is associated with the Lie group $SO(2M)$, but adds an extra term which leads to parity violation.  This approach can be understood\cite{Fukutome1977b,Nishiyama2019} as an embedding of $SO(2M+1)$ into $SO(2M+2)$; the former is the appropriate Lie group when number parity violations are included.  In 2018, Moussa proposed a similar idea\cite{Moussa2018} which extends the generators of $SO(2M)$ with single fermion operators times the parity operator to construct $SO(2M+1)$ and then generates a canonical transformation.  A third possibility is textbook material \cite{Blaizot1985}, and consists of adding a shift term of Grassmann numbers to HFB unitary $SO(2M)$ canonical rotations.  We are unaware of any implementation of these methods, and the development of the theory is not entirely straightforward.

In this manuscript, we thus have two main goals. First, we give a simplified presentation of the main ideas behind these exotic mean-field states and discuss a 4th alternative where we construct all Fock states of even and odd parity using the coherent states of $SO(2M+1)$ \cite{Perelomov1986}.  Although this simplified picture has been mentioned in the literature \cite{Fukutome1977b,Dobaczewski1982} as a possibility, it was not elaborated and never implemented.  We show that this approach is equivalent to all these previous formulations. Second, we show how these mean-field methods can be used to solve Jordan-Wigner–transformed spin systems, which yield fermionic Hamiltonians that do not conserve number parity. As our benchmarks show, for frustrated spin systems, doing the JW transformation in a basis where $S_z$ is broken yields results superior to those obtained with number-parity-conserving Hamiltonians.

\section{Number-Parity--Violating Mean-Field}
The number-parity--violating mean-field theories we have mentioned are cast as canonical transformations of the underlying fermionic creation and annihilation operators.  To understand how these theories work and why they are developed the way they are, we start by looking at bosonic mean-field theory, for which number-parity-violating mean-field theories are simple.

\subsection{Canonical Transformations for Bosonic Operators}
Suppose we have a set of bosonic operators $(b^\dag_p, b_q)$, for $1 \leq p,\, q \leq M$, with canonical commutation relations
\begin{subequations}
\begin{align}
[b_p^\dag,b_q^\dag] &= 0 = [b_p,b_q],
\\
[b_p,b_q^\dagger] &= \delta_{pq}.
\end{align}
\end{subequations}

We can consider a simple linear transformation which mixes the various creation operators:\cite{Blaizot1985}
\begin{equation}
\tilde{b}_p^\dag = U_{ip} \, b_i^\dag,
\end{equation}
where $\boldsymbol{U}$ is a complex matrix of parameters and where we have assumed, as we will throughout this work, that repeated indices are summed.  If the transformation is to be canonical, the new operators $\tilde{b}^\dagger$ and $\tilde{b}$ must obey the same commutation relations as do $b^\dagger$ and $b$.  If we insert our expression for $\tilde{b}$ and its adjoint into the commutators, we find that
\begin{subequations}
\begin{align}
[\tilde{b}_p^\dag,\tilde{b}_q^\dag] &= U_{ip} \, U_{jq} \, [b_i^\dag,b_j^\dag] = 0,
\\
[\tilde{b}_p,\tilde{b}_q^\dagger] &= U_{ip}^\star \, U_{jq} \, [b_i,b_j^\dagger] = U_{ip}^\star \, U_{iq} = \left(\boldsymbol{U}^\dag \, \boldsymbol{U}\right)_{pq}.
\end{align}
\end{subequations}
Thus, we see that requiring the transformation to be canonical means that we need $\boldsymbol{U}^\dag \, \boldsymbol{U} = \boldsymbol{1}$, i.e. the matrix $\boldsymbol{U}$ must be unitary.  Such transformations are used, for example, in Hartree-Bose theory.

We could try a more general transformation which mixes creation and annihilation operators, as we would use in Hartree-Bose-Bogoliubov:
\begin{equation}
\beta_p^\dag = U_{ip} \, b_i^\dag + V_{ip} \, b_i.
\end{equation}
Repeating the same procedure of insisting that $\beta^\dag$ and $\beta$ satisfy bosonic commutation relations, we arrive at a pair of constraints on $\boldsymbol{U}$ and $\boldsymbol{V}$:
\begin{subequations}
\label{Eqn:HBBConditions}
\begin{align}
\boldsymbol{U}^\T \, \boldsymbol{V} - \boldsymbol{V}^\T \, \boldsymbol{U} &= \boldsymbol{0},
\\
\boldsymbol{U}^\dag \, \boldsymbol{U} - \boldsymbol{V}^\dag \, \boldsymbol{V} &= \boldsymbol{1}.
\end{align}
\end{subequations}
The first of these comes from commutation of $\beta_p^\dag$ with $\beta_q^\dag$, and the second of $\beta_p$ with $\beta_q^\dag$.  The relative minus sign between the two terms is a consequence of bosonic operators using the commutator, so that $[b,b^\dagger] = -[b^\dagger,b]$.  Noting that the transformation of $b$ and $b^\dagger$ can be written as
\begin{equation}
\begin{pmatrix}	\beta_p^\dag & \beta_q	\end{pmatrix}
=
\begin{pmatrix}	b_i^\dag & b_j	\end{pmatrix} \,
\begin{pmatrix}
    U_{ip} & V_{iq}^\star \\
    V_{jp} & U_{jq}^\star
\end{pmatrix},
\end{equation}
we can express the canonical commutation conditions of Eqn. \ref{Eqn:HBBConditions} in terms of constraints on the matrix ${\boldsymbol{W} = \left( \begin{smallmatrix}\boldsymbol{U} & \boldsymbol{V}^\star \\ \boldsymbol{V} & \boldsymbol{U}^\star\end{smallmatrix}\right)}$, as
\begin{equation}
\boldsymbol{W}^\dag \, \boldsymbol{\eta} \, \boldsymbol{W} = \boldsymbol{\eta},
\end{equation}
where $\boldsymbol{\eta} = \left(\begin{smallmatrix} \boldsymbol{1} & \boldsymbol{0} \\ \boldsymbol{0} & -\boldsymbol{1}\end{smallmatrix}\right)$.  That this is precisely the orthonormalization condition for the eigenvectors in the random phase approximation is not an accident.\cite{Blaizot1985}

So far, we have seen bosonic transformations to a set of operators $\tilde{b}^\dag$ and $\tilde{b}$ where the transformation conserves number (that is, $\tilde{b}_p^\dag$ acts on a number eigenstate to produce another number eigenstate), and another to a set of operators $\beta^\dag$ where the transformation does not conserve number but does conserve number parity (because $\beta_p^\dag$ acts on a number parity eigenstate to produce another number parity eigenstate).  To break number parity, we can go one step further, and define
\begin{equation}
B_p^\dag = U_{ip} \, b_i^\dag + V_{ip} \, b_i + y_p^\star,
\end{equation}
where $y_p$ is a complex number which we will refer to as the complex shift.  Because complex numbers commute with other complex numbers and with bosonic operators, the conditions on $\boldsymbol{U}$ and $\boldsymbol{V}$ do not change, but now the operator $B_p$ acts on a state of definite number parity to return a state of mixed number parity.  So for bosons, all of this is straightforward.

\subsection{Canonical Transformations for Fermionic Operators}
Now let us repeat the process for fermions.  We have a set of fermionic operators $(a_p^\dag,a_q)$, for $1 \leq p,\, q \leq M$, which obey canonical anticommutation relations
\begin{subequations}
\begin{align}
\{a_p^\dag,a_q^\dag\} &= 0 = \{a_p,a_q\},
\\
\{a_p,a_q^\dagger \} &= \delta_{pq}.
\end{align}
\end{subequations}
We can mix the operators in a number-conserving way, as we do in Hartree-Fock:
\begin{equation}
\tilde{a}_p^\dag = U_{ip} \, a_i^\dag.
\end{equation}
Insisting that the transformation is canonical just means that $\boldsymbol{U}$ must be unitary.  We can mix the operators in a number-nonconserving way, as we do in Hartree-Fock-Bogoliubov:
\begin{equation}
\label{Eqn:DefAlpha}
\alpha_p^\dag = U_{ip} \, a_i^\dag + V_{ip} \, a_i.
\end{equation}
Insisting that the transformation be canonical means that the matrix ${\boldsymbol{W} = \left( \begin{smallmatrix}\boldsymbol{U} & \boldsymbol{V}^\star \\ \boldsymbol{V} & \boldsymbol{U}^\star\end{smallmatrix}\right)}$ must be unitary.

If we try to break number parity in a way analogous to what is done for bosons, by defining
\begin{equation}
A_p^\dag = U_{ip} \, a_i^\dag + V_{ip} \, a_i + y_p^\star = \alpha_p^\dag + y_p^\star,
\end{equation}
for complex $y_p$, however, we run into a problem: the constraints we must impose become
\begin{subequations}
\begin{align}
\{A_p^\dag,A_q^\dag\} &= \{\alpha_p^\dag,\alpha_q^\dag\} + 2 \, y_p^\star \, \alpha_q^\dag + 2 \, y_q^\star \, \alpha_p^\dag + 2 \, y_p^\star \, y_q^\star \\ & = 0 \nonumber,
\\
\{A_p,A_q^\dagger\} &= \{\alpha_p,\alpha_q^\dagger\} + 2 \, y_p \, \alpha_q^\dagger + 2 \, y_q^\star \, \alpha_p + 2 \, y_p \, y_q^\star  \\ &= \delta_{pq}. \nonumber
\end{align}
\end{subequations}
This we cannot do: the anticommutators $\{\alpha_p^\dag,\alpha_q^\dag\}$ and $\{\alpha_p,\alpha_q^\dagger\}$ are just complex numbers which depend on the coefficients $\boldsymbol{U}$ and $\boldsymbol{V}$ and we cannot use them to cancel operator terms like $2 \, y_p^\star \, \alpha_q^\dag + 2 \, y_q^\star \, \alpha_p^\dag$.

One way to solve the problem\cite{Blaizot1985} is to choose the shifts $\boldsymbol{y}$ to be Grassmann numbers such that they anticommute with each other and with the fermionic operators $a$ and $a^\dagger$.  If we do so, then insisting that the transformation from $(a^\dag, a)$ to $(A^\dag, A)$ must be canonical means that $\boldsymbol{W}$ must be unitary.  The price we must pay is that we have to work with Grassmann numbers, which are not very familiar to most of us and for which there is limited support for numerical linear algebra.

Alternatively, we can try to modify the transformation to define $A_p^\dag$ such that we still use complex numbers, with operators which generate the minus signs we need to convert the anticommutators $\{y_p,\alpha^\dag_q\}$ and $\{y^\star_p, \alpha_q\}$ into commutators, which vanish.  If we can do this, then we can hope for nonlinear canonical transformations of fermionic operators with only complex parameters.

\subsection{The Fukutome and Moussa Constructions}
Suppose that we can find operators $P$ and $Q$ such that
\begin{subequations}
\label{Eqn:ShiftedFermionicTransformation}
\begin{align}
F_p^\dagger &= P^\dagger \, \left(\alpha_p^\dagger + \frac{1}{\sqrt{2}} \, y_p^\star\right) = \left(\alpha_p^\dagger - \frac{1}{\sqrt{2}} \, y_p^\star\right) \, Q^\dagger,
\\
F_p &= \left(\alpha_p + \frac{1}{\sqrt{2}} \, y_p\right) \, P = Q \, \left(\alpha_p - \frac{1}{\sqrt{2}} \, y_p\right),
\end{align}
\end{subequations}
and such that $P \, Q = \lambda$ for some nonzero complex number $\lambda$ while $P \, P^\dagger = Q^\dagger \, Q = \Lambda$ for some nonzero real number $\Lambda$.  The factors of $1/\sqrt{2}$ are for later convenience, and the operator $\alpha_p^\dag$ is a quasiparticle annihilation operator as defined in Eqn. \ref{Eqn:DefAlpha} except that the matrix $\boldsymbol{W}$ need not be unitary.  This is because the condition that $\boldsymbol{W}$ was unitary enforced that the quasiparticle operators $\alpha^\dag$ and $\alpha$ obeyed canonical anticommutation relations.  Here, instead, we want $F^\dag$ and $F$ to do so.  To help guide the reader through the presentation, Table \ref{Table:Dictionary} defines the most important symbols we will need in constructing these shifted quasiparticle operators.

\begin{table}[t]
\caption{Dictionary of important symbols.\label{Table:Dictionary}}
\begin{ruledtabular}
\begin{tabular}{cl}
Symbol	&	Meaning	\\
\hline
$a^\dagger$, $a$ 				&	Bare fermion creation and annihilation operators	\\
$\alpha^\dagger$, $\alpha$ 		&	Bogoliubov creation and annihilation operators	\\
$F^\dagger$, $F$				&	Fukutome creation and annihilation operators	\\
$P$, $Q$					&	Operators used to help define $F$ and $F^\dagger$	\\
$\gamma$					&	Operator used to define $P$ and $Q$	\\
$\boldsymbol{U}$, $\boldsymbol{V}$	&	Coefficients relating $(a^\dagger, a)$ to $(\alpha^\dagger, \alpha)$ or $(F^\dagger, F)$\\
$\boldsymbol{y}$				&	Vector of complex shifts	\\
$\boldsymbol{x}$				&	Vector of parameters in the operator $\gamma$	\\
$z$						&	Scalar coefficient in $P$ and $Q$	\\
$\boldsymbol{G}$				&	Unitary matrix collecting $\boldsymbol{U}$, $\boldsymbol{V}$, $\boldsymbol{x}$, $\boldsymbol{y}$, and $z$	\\
$Z$						&	Thouless transformation operator with coefficients $\boldsymbol{Z}$	\\
$t^\dagger$					&	Blocked level creation operator with coefficients $\boldsymbol{t}$	\\
\end{tabular}
\end{ruledtabular}
\end{table}

The anticommutators we need are
\begin{subequations}
\begin{align}
\{F_p^\dag,F_q^\dag\}
	&=	\left(\alpha_p^\dagger - \frac{1}{\sqrt{2}} \, y_p^\star\right) \, Q^\dag \, P^\dag \, \left(\alpha_q^\dagger + \frac{1}{\sqrt{2}} \, y_q^\star\right)
\\
	&+ \left(\alpha_q^\dagger - \frac{1}{\sqrt{2}} \, y_q^\star\right) \, Q^\dag \, P^\dag \, \left(\alpha_p^\dagger + \frac{1}{\sqrt{2}} \, y_p^\star\right)
\nonumber
\\
	&=	\lambda^\star \, \left(\{\alpha_p^\dagger,\alpha_q^\dagger\} - y_p^\star \, y_q^\star\right),
\nonumber
\\
\{F_p,F_q^\dagger\}
	&=	\left(\alpha_p + \frac{1}{\sqrt{2}} \, y_p\right) \, P \, P^\dagger \, \left(\alpha_q^\dagger + \frac{1}{\sqrt{2}} \, y_q^\star\right)
\\
	&+	\left(\alpha_q^\dagger - \frac{1}{\sqrt{2}} \, y_q^\star\right) \, Q^\dagger \, Q \, \left(\alpha_p- \frac{1}{\sqrt{2}} \, y_p\right)
\nonumber
\\
	&=	\Lambda \, \left(\{\alpha_p,\alpha_q^\dagger\} + y_p \, y_q^\star\right).
\nonumber
\end{align}
\end{subequations}
Recalling that
\begin{equation}
  \begin{pmatrix}
    \alpha_p^\dag & \alpha_q
  \end{pmatrix}
  = \begin{pmatrix}
    a_i^\dag & a_j
  \end{pmatrix} \,
  \begin{pmatrix}
    U_{ip} & V_{iq}^\star \\
    V_{jp} & U_{jq}^\star
  \end{pmatrix},
\end{equation}
we have
\begin{subequations}
\begin{align}
\{\alpha_p^\dag,\alpha_q^\dag\} &= U_{ip} \, V_{iq} + V_{ip} \, U_{iq},
\\
\{\alpha_p,\alpha_q^\dagger\} &= U_{ip}^\star \, U_{iq} + V_{ip}^\star \, V_{iq}.
\end{align}
\end{subequations}
Then for the transformation of Eqn. \ref{Eqn:ShiftedFermionicTransformation} to be canonical, we simply require
\begin{subequations}
\label{Eqn:UVConstraints}
\begin{align}
0 &= U_{ip} \, V_{iq} + V_{ip} \, U_{iq} - y_p^\star \, y_q^\star
\\
  &= \left(\boldsymbol{U}^\T \, \boldsymbol{V} + \boldsymbol{V}^\T \, \boldsymbol{U}  - \boldsymbol{y}^\star \, \boldsymbol{y}^\dag\right)_{pq},
\nonumber
\\
\delta_{pq} &= \Lambda \, \left(U_{ip}^\star \, U_{iq} + V_{ip}^\star \, V_{iq} + y_p \, y_q^\star\right)
\label{Eqn:DpqConstraint}
\\
 &= \Lambda \, \left(\boldsymbol{U}^\dag \, \boldsymbol{U} + \boldsymbol{V}^\dag \, \boldsymbol{V} + \boldsymbol{y} \, \boldsymbol{y}^\dagger\right)_{pq},
\nonumber
\end{align}
\end{subequations}
where $\boldsymbol{y}$ is the column vector with entries $y_p$.  If we can somehow find the necessary operators $P$ and $Q$ such that
\begin{subequations}
\label{Eqn:ABConstraints}
\begin{align}
\left(\alpha_p + \frac{1}{\sqrt{2}} \, y_p\right) \, P &= Q \, \left(\alpha_p - \frac{1}{\sqrt{2}} \, y_p\right),
\\
P \, Q &= \lambda,
\\
P \, P^\dagger &= Q^\dagger \, Q = \Lambda,
\end{align}
\end{subequations}
and if we can satisfy the constraints of Eqn. \ref{Eqn:UVConstraints}, then we have successfully generated a nonlinear number-parity--violating fermionic canonical transformation.  Fukutome and Moussa propose two different approaches (i.e. two different choices for $P$ and $Q$), but both fall within the same basic framework.

In Fukutome's case \cite{Fukutome1977,Fukutome1977b}, the operators $P$ and $Q$ are given by
\begin{subequations}
\label{Eqn:DefPQ}
\begin{align}
P &= z + \gamma,
\\
Q &= z - \gamma,
\\
\gamma &= \sqrt{2} \, x_p^\star \, a_p - \sqrt{2} \, x_p \, a_p^\dagger,
\end{align}
\end{subequations}
where $z$ is a real number.  Note that $\gamma^\dagger = -\gamma$ which means $Q = P^\dagger$, and that $\gamma^2 = -2 \, x_p \, x_p^\star$, which follows from the canonical anticommutation relations of the basic fermionic operators $a^\dag$ and $a$.  Enforcing the constraints of Eqn \ref{Eqn:ABConstraints} requires
\begin{subequations}
\label{Eqn:PQConstraintsFukutome}
\begin{align}
  U_{ip}^\star \, x_i - V_{ip}^\star \, x_i^\star - z \, y_p &= 0,
  \\
  2 \, x_p^\star \, x_p + z^2 &= \Lambda.
\end{align}
\end{subequations}
We may as well follow Fukutome's choice of $\Lambda = 1$.

In Moussa's case \cite{Moussa2018}, the operators are instead
\begin{equation}
P = \Pi = -Q.
\end{equation}
Because $\Pi = \Pi^\dagger$, $\Pi^2 = 1$ and $\{\Pi,a_p\} = 0$, the constraints of Eqn. \ref{Eqn:ABConstraints} are all satisfied automatically, with $\lambda = -1$ and $\Lambda = 1$.

In both cases, it remains to enforce the canonical anticommutation constraints of Eqn. \ref{Eqn:UVConstraints}.  We will specialize to the case $\Lambda = 1$ (other values can be absorbed into scaling $P$ and $Q$).  Then consider the matrix
\begin{equation}
\boldsymbol{G} = 
\begin{pmatrix}
\boldsymbol{U} & \boldsymbol{V}^\star & \boldsymbol{r} \\
\boldsymbol{V} & \boldsymbol{U}^\star & \boldsymbol{s} \\
\boldsymbol{y}^\dag & -\boldsymbol{y}^\T & \theta
\end{pmatrix}
\end{equation}
where $\boldsymbol{r}$ and $\boldsymbol{s}$ are column vectors to which we shall return presently, while $\theta$ is just a number.  Insisting that $\boldsymbol{G}$ is unitary gives us a list of equations:
\begin{subequations}
\label{Eqns:GUnitary}
\begin{align}
  \boldsymbol{U}^\dag \, \boldsymbol{U} + \boldsymbol{V}^\dag \, \boldsymbol{V} + \boldsymbol{y} \, \boldsymbol{y}^\dag
  &= \mathbf{1},
  \\
  \boldsymbol{U}^\T \, \boldsymbol{V} + \boldsymbol{V}^\T \, \boldsymbol{U}  - \boldsymbol{y}^\star \, \boldsymbol{y}^\dag
  &= \mathbf{0},
  \\
  \boldsymbol{U}^\dag \, \boldsymbol{r} + \boldsymbol{V}^\dag \, \boldsymbol{s} + \theta \, \boldsymbol{y}
  &= \boldsymbol{0},
  \\
  \boldsymbol{V}^\T \, \boldsymbol{r} + \boldsymbol{U}^\T \, \boldsymbol{s} - \theta \, \boldsymbol{y}^\star
  &= \boldsymbol{0},
  \\
  \boldsymbol{r}^\dagger \, \boldsymbol{r} + \boldsymbol{s}^\dagger \, \boldsymbol{s} + \theta^\star \, \theta
  &= 1.
\end{align}
\end{subequations}
The other four blocks of the matrix equation $\boldsymbol{G}^\dag \, \boldsymbol{G} = \boldsymbol{1}$ are complex conjugates or adjoints of equations already provided.

Clearly, unitarity of $\boldsymbol{G}$ is sufficient to satisfy the constraints of Eqns. \ref{Eqn:UVConstraints}.  Choosing $\boldsymbol{r}$, $\boldsymbol{s}$, and $\theta$ to be $-\boldsymbol{x}$, $\boldsymbol{x}^\star$, and $z$, respectively, in Fukutome's formulation gives us the remaining constraints of Eqn. \ref{Eqn:PQConstraintsFukutome}.

Note finally that the matrix $\boldsymbol{G}$ in Fukutome's formulation is a representation of the $SO(2M+1)$ Lie group, which is sensible as the $so(2M+1)$ Lie algebra can be generated by $a^\dag_p$, $a_p$, $a_p^\dag \, a_q^\dag$, $a_p \, a_q$, and $a_p^\dag \, a_q - \frac{1}{2}\, \delta_{pq}$.\cite{Wybourne1973, Fukutome1977}  To see this, we note that we can transform $\boldsymbol{G}$ to be real (hence orthogonal, since it is already unitary) by
\begin{align}
\nonumber
\boldsymbol{O}
 &= \boldsymbol{A} \, \boldsymbol{G} \, \boldsymbol{A}^{-1}
\\
 &= \begin{pmatrix}
   \mathrm{Re}(\boldsymbol{U} + \boldsymbol{V}) & \mathrm{Im}(\boldsymbol{U} + \boldsymbol{V}) & -\sqrt{2}\, \mathrm{Im}(\boldsymbol{x})\\
  -\mathrm{Im}(\boldsymbol{U} - \boldsymbol{V}) & \mathrm{Re}(\boldsymbol{U} - \boldsymbol{V}) & -\sqrt{2}\, \mathrm{Re}(\boldsymbol{x})\\
  -\sqrt{2} \, \mathrm{Im}(\boldsymbol{y})^\T   & -\sqrt{2} \, \mathrm{Re}(\boldsymbol{y})^\T  & z
	\end{pmatrix},
\end{align}
where
\begin{equation}
\boldsymbol{A} = \begin{pmatrix} \frac{\mathbf{i}}{\sqrt{2}}  & \frac{\mathbf{i}}{\sqrt{2}} & \boldsymbol{0} \\ -\frac{\boldsymbol{1}}{\sqrt{2}} & \frac{\boldsymbol{1}}{\sqrt{2}} & \boldsymbol{0} \\ \boldsymbol{0} & \boldsymbol{0} & 1 \end{pmatrix}.
\end{equation}
We can always choose the sign of $z$ such that $\boldsymbol{O}$, and hence $\boldsymbol{G}$, has determinant 1 (see Appendix~\ref{Sec:FukutomeProof}).

Once the canonical transformation is established, one can develop a mean-field theory which, in analogy with HFB, we shall refer to as Hartree-Fock-Bogoliubov-Fukutome (HFBF).  For details, we refer the reader to Fukutome's 1977 manuscript (Ref. \onlinecite{Fukutome1977}).  Those details will not be required here.  Suffice it to say that we define quasiparticle operators $F$ and $F^\dagger$ in terms of which we build the Fukutome vacuum $|0_F\rangle = \left(\prod_i F_i\right) |-\rangle$ where $|-\rangle$ is the physical vacuum.  We then define the energy as the expectation value of the Hamiltonian with respect to $|0_F\rangle$, possibly with the inclusion of a chemical potential term $\mu \, N$ to control the average number of physical particles.   The energy is then minimized with respect to the parameters defining the transformation matrix $\boldsymbol{G}$.  Evaluating the energy is straightforward given the density matrices, which we briefly discuss in Appendix \ref{Sec:DensityMatrices}.

\section{A Simplified Picture}
So far, we have discussed everything from the rarefied perspective of canonical transformations.  The physical interpretation of this approach is not entirely straightforward, and we feel it might be easier to consider an alternative but equivalent formulation.\cite{Fukutome1977b}

Consider, then, a Hartree-Fock-Bogoliubov state.  Generically, we may parametrize these number-violating mean-field states in terms of a Thouless transformation\cite{Thouless1960}
\begin{equation}
|\Phi(Z)\rangle = \mathrm{e}^Z |0\rangle
\label{Def:PhiZ}
\end{equation}
where
\begin{equation}
Z = \sum_{p>q} Z_{pq} \, \alpha_p^\dagger \, \alpha_q^\dagger
\end{equation}
and where $|0\rangle$ is some reference Bogoliubov determinant with good number parity, with respect to which the quasiparticle creation operators $\alpha_p^\dagger$ are defined, i.e. $|0\rangle$ is the vacuum for $\alpha$.  When the Bogoliubov reference has even number parity -- and in this case it may generally be chosen to be the physical vacuum -- we have an even number-parity HFB state.  When the reference has odd number parity such that a level is blocked (i.e. is singly-occupied as opposed to being in a linear combination of empty and double occupancy), we have an odd number-parity HFB state.\cite{Ring1980}

The simplest way to generalize this construction to have sectors of both even and odd number parity is to modify the reference state $|0\rangle$.  Specifically, we wish to generalize the Bogoliubov state in the following straightforward way:
\begin{equation}
|\Phi(Z,t)\rangle = \mathrm{e}^Z \, \left(1 +  t_p \, \alpha_p^\dagger\right) |0\rangle
\label{Eqn:WFN1}
\end{equation}
where $|0\rangle$ is an even-parity HFB state.  That is, our mean-field state is a linear combination of an even number-parity Bogoliubov state, defined by the Thouless transformation $Z$ and the reference $|0\rangle$, and an odd number-parity Bogoliubov state which is defined by the same Thouless transformation $Z$ acting on an odd-parity reference which we create by blocking the level with the operator
\begin{equation}
t^\dagger =  t_p \, \alpha_p^\dagger.
\end{equation}
It turns out that this simple state is exactly the vacuum of Moussa's mean-field\cite{Moussa2018}, implemented as a non-unitary rather than a unitary Thouless transformation.  In Appendix \ref{Sec:FukutomeProof} we show that it is also equivalent to Fukutome's mean-field.

Note that, because $t^\dagger$ is a (non-normalized) fermionic creation operator, it is nilpotent.  This means that
\begin{equation}
1 + t^\dagger = \mathrm{e}^{t^\dagger}.
\end{equation}
Moreover, $t^\dagger$ commutes with $Z$, so our mean-field state can be written as
\begin{equation}
|\Phi(Z,t)\rangle = \mathrm{e}^{Z + t^\dagger} |0\rangle.
\end{equation}
This wave function is a coherent state of $SO(2M+1)$  \cite{Dobaczewski1982} defined by exponentiating the corresponding Lie algebra generators, in the same we we build HF as $U(M)$ and HFB as $SO(2M)$ coherent states.\cite{Blaizot1985,Perelomov1986}   Finally, we may move $t^\dagger$ to the left so that
\begin{equation}
|\Phi(Z,t)\rangle = \left(1 + t^\dagger\right) \, \mathrm{e}^Z |0\rangle = \left(1 + t^\dagger\right) |\Phi(Z)\rangle
\end{equation}
where $|\Phi(Z)\rangle$ is an even-parity Bogoliubov state.  We may then transform the quasiparticle creation operators $\alpha^\dagger$ in defining $t^\dagger$ into quasiparticle creation and annihilation operators acting on $|\Phi(Z)\rangle$:
\begin{equation}
t_p \, \alpha_p^\dagger =  \tilde{t}_p(Z) \, \alpha_p^\dagger(Z) +  \bar{t}_p(Z) \, \alpha_p(Z)
\end{equation}
where $\alpha_p(Z)$ annihilates $|\Phi(Z)\rangle$ and where $\tilde{\boldsymbol{t}}$ and $\bar{\boldsymbol{t}}$ are the original amplitudes $\boldsymbol{t}$ contracted with the $\boldsymbol{U}$ and $\boldsymbol{V}$ amplitudes associated with the Thouless transformation $\mathrm{e}^Z$.  We can drop the $\alpha_p(Z)$ term, and if we wish we may write simply
\begin{equation}
|\Phi(Z,t)\rangle = \left(1 + t_p \, \alpha_p^\dagger(Z)\right) \, |\Phi(Z)\rangle
\label{Eqn:WFN2}
\end{equation}
in a slight abuse of notation since the amplitudes $\boldsymbol{t}$ in Eqns. \ref{Eqn:WFN1} and \ref{Eqn:WFN2} are different.

The advantage of working with the wave function in the form given by Eqn. \ref{Eqn:WFN2} is simply that it facilitates computation.  While density matrices can be derived from Fukutome's procedure and he presents expressions for them, we find it conceptually simpler to understand everything as a kind of half-body configuration interaction and parametrize everything in terms of the HFB state $|\Phi(Z)\rangle$ associated with our parity-broken mean-field.  Thus, for example, we can evaluate the overlap as
\begin{align}
\langle \Phi(Z,t)&|\Phi(Z,t)\rangle
\nonumber
\\
 =& \langle \Phi(Z) | \left(1 + t\right) \, \left(1 + t^\dagger\right) |\Phi(Z)\rangle
\nonumber
\\
 =& \langle \Phi(Z)| 1 + t + t^\dagger + t^\dagger \, t + \{t^\dagger,t\} |\Phi(Z)\rangle
\nonumber
\\
 =& \left(1 + t_p^\star \, t_p\right) \, \langle \Phi(Z)|\Phi(Z)\rangle
\end{align}
where we have used that $t$ annihilates $|\Phi(Z)\rangle$ and $t^\dagger$ annihilates $\langle \Phi(Z)|$, while
\begin{equation}
\{t,t^\dagger\} = t_p^\star \, t_q \, \{\alpha_p,\alpha_q^\dagger\} = t_p^\star \, t_p.
\end{equation}

\section{Number Parity Violation in Jordan-Wigner--Transformed Hamiltonians}
Now that we have introduced the central ideas behind number-parity--violating fermionic mean-fields, let us turn our attention to the Jordan-Wigner transformation.

Introduced at the dawn of quantum mechanics\cite{Jordan1928} , the Jordan-Wigner (JW) transformation establishes a duality, an equivalence between systems of spin 1/2 and systems of spinless fermions, by writing
\begin{subequations}
\label{Eqn:DefJW}
\begin{align}
S_p^+ &\mapsto a_p^\dagger \, \phi_p^\dagger,
\\
S_p^- &\mapsto a_p \, \phi_p,
\\
S_p^z &\mapsto \bar{n}_p = n_p - \frac{1}{2},
\\
\phi_p^\dagger &= \phi_p = \mathrm{e}^{\mathrm{i} \, \pi \, \sum_{k<p} n_k} = \prod_{k<p} \left(1 - 2 \, n_k\right),
\label{Eqn:JWStrings}
\\
n_p &= a_p^\dagger \, a_p,
\end{align}
\end{subequations}
It is textbook material\cite{Nishimori2011} that one can take advantage of the JW transformation to convert the XXZ Heisenberg Hamiltonian
\begin{equation}
H_\mathrm{XXZ} = \sum_{\langle pq \rangle} \left[\frac{1}{2} \, \left(S_p^+ \, S_q^- + S_p^- \, S_q^+\right) + \Delta \, S_p^z \, S_q^z\right]
\end{equation}
into a fermionic Hamiltonian 
\begin{align}
H_\mathrm{XXZ}^\mathrm{JW} = \sum_{\langle pq \rangle} \Big[&\frac{1}{2} \, \left(a_p^\dagger \, \phi_p \, \phi_q \, a_q + a_p \, \phi_p \, \phi_q \, a_q^\dagger\right)
\\
 &+ \Delta \, \left(n_p - \frac{1}{2}\right) \, \left(n_q - \frac{1}{2}\right)\Big]
\nonumber
\end{align}
which, at $\Delta = 0$, is non-interacting, where we have used the fact that the Hamiltonian only couples neighboring sites $p$ and $q$ (as indicated by the symbol $\langle pq \rangle$) for which the JW string operators $\phi_p$ and $\phi_q$ vanish.  Note that at $\Delta = 0$, the exact wave function in terms of spins is rather complicated, being of the form
\begin{equation}
|\Psi_\mathrm{XXZ}\rangle_{\Delta = 0} = \exp(\sum_{p>q} J_{pq} \, S_p^z \, S_q^z) \, \left(\sum_r \eta_r \, S_r^+\right)^n \lvert \Downarrow\rangle
\end{equation}
where $n$ is the number of $\uparrow$ spins in the system and $\lvert\Downarrow\rangle$ is a product state in which each spin points in the $\downarrow$ direction.\cite{Liu2023}  Yet for all its complexity, this correlated spin state is equivalent to a simple mean-field state in the JW-transformed picture.

Inspired by this result, and taking advantage of the fact that the JW strings are exponentials of one-body operators and thus Thouless transformations, we investigated using the JW transformation to convert a handful of spin systems to fermionic systems which we can then solve at the mean-field level.\cite{Henderson2022,Chen2023,Henderson2023}  Our initial work faced an important limitation, however: the kinds of spin Hamiltonians we could treat were sharply limited by the need to avoid number-parity violating terms.  These would naturally arise from terms such as $S_p^+ \, S_q^z$, for example, which would map to $a_p^\dagger \, \phi_p \, \left(n_q - 1/2\right)$.  Our work here shows that Hamiltonians with such terms can be handled by fermionic mean-field methods nonetheless.  One simply needs a more sophisticated fermionic mean-field.

\subsection{Noninteracting Hamiltonians}
Let us begin with a simple non-interacting spin Hamiltonian
\begin{equation}
H = \sum \vec{J}_p \cdot \vec{S}_p
\end{equation}
where $\vec{J}_p$ is a vector of real coefficients and $\vec{S}_p$ is the spin vector $(S_p^x,S_p^y,S_p^z)$. It has been pointed out\cite{Ryabinkin2018b} that such Hamiltonians can be exactly solved by a spin mean-field procedure but their JW-transformed versions cannot be solved via Hartree-Fock.  This is unquestionably true, but we show here that they can be solved by this more general mean-field procedure.

To see this, let us start by looking at the solution of the original Hamiltonian.  Because the Hamiltonian is non-interacting, its ground state wave function is quite simple:
\begin{equation}
|\Psi\rangle = \prod_i \left(u_i + v_i \, S_i^+\right) \lvert \Downarrow \rangle
\end{equation}
where $\lvert\Downarrow\rangle$ is the spin vacuum in which each spin is an $S^z$ eigenstate with eigenvalue $-1/2$.   In other words, the wave function is a spin version of the Bardeen-Cooper-Schrieffer wave function.  We will assume that the parameters $u_i$ are non-zero, which we can contrive by ensuring that none of the $\vec{J}_i$ point strictly along the $z$ axis, in which case we can simply write
\begin{equation}
|\Psi\rangle = \prod_i \left(1 + \eta_i \, S_i^+\right) \lvert \Downarrow\rangle
\end{equation}
for complex parameters $\eta$.  When $u_i$ does vanish, the site $i$ is strictly spin up, which corresponds to a fermionic orbital which is fully occupied and a spectator to the following analysis.

Upon JW transformation, the spin vacuum maps to the physical vacuum, denoted here by $|-\rangle$.  The JW transformation of the ground state wave function is then
\begin{equation}
|\Psi\rangle \mapsto |\Psi_\mathrm{JW}\rangle = \prod_i \left(1 + \eta_i \, a_i^\dagger \, \phi_i\right) |-\rangle.
\end{equation}
In the spin case, the various spin operators $S_i^+$ all commute so it does not matter in which order we understand the product.  The JW strings ensure that the same is true after JW transformation.  In particular, we can choose to understand the product so that $a_i^\dagger \, \phi_i$ appears to the left of $a_j^\dagger \, \phi_j$ when $j > i$.  That being the case, the JW strings $\phi_i$ commute all the way to the right (because $\phi_i$ commutes with $a_j^\dagger$ for $j > i$) where they reach the physical vacuum.  Since $\phi_i |-\rangle = |-\rangle$, that means the mapped wave function is
\begin{equation}
|\Psi\rangle \mapsto |\Psi_\mathrm{JW}\rangle = \prod_i \left(1 + \eta_i \, a_i^\dagger\right) |-\rangle
\end{equation}
with the product understood to be in ascending order from left to right.  Such a state is of the form
\begin{equation}
|\Psi_{JW}\rangle = \prod_k F_k |-\rangle,
\end{equation}
as is easiest to see by writing $F_k = \left(\eta_k \, a_k^\dagger + (-1)^{k-1}\right) \, \Pi$ and moving the parity operators through to reach the physical vacuum.  The sign originates because the operator $F_k$ has $k-1$ parity operators $\Pi$ behind it which anticommute with $a_k^\dagger$ to give a factor of $(-1)^{k-1}$.  Operators of this form are vacuums for a Fukutome-style state, just as a standard HFB vacuum is given by acting all quasiparticle annihilation operators on the physical vaccum.\cite{Ring1980}  All of this is to say that the JW-transformation of the exact ground-state wave function for non-interacting spin Hamiltonians is of this general number-parity--violating mean-field form.

We note in passing that we can equivalently write the state as
\begin{subequations}
\begin{align}
|\Psi_\mathrm{JW}\rangle &= \left(1 + t^\dagger\right) \, \mathrm{e}^Z |-\rangle,
\\
t^\dagger &= \sum \eta_p \, a_p^\dagger,
\\
Z &= \sum_{p<q} \eta_p \, \eta_q \, a_p^\dagger \, a_q^\dagger,
\end{align}
\end{subequations}
which is a special case of the wave function given in Eqn. \ref{Eqn:WFN1}.  Without going into too much detail, observe that 
\begin{equation}
\frac{1}{n!} \, Z^n = \sum_{p_1 < \ldots < p_{2n}} \eta_{p_1} \ldots \eta_{p_{2n}} \, a_{p_1}^\dagger \ldots a_{p_{2n}}^\dagger
\end{equation}
and that
\begin{equation}
\frac{1}{n!} \, t^\dagger \, Z^n = \sum_{p_1 < \ldots < p_{2n+1}} \eta_{p_1} \ldots \eta_{p_{2n+1}} \, a_{p_1}^\dagger \ldots a_{p_{2n+1}}^\dagger
\end{equation}
which is what we get by expanding the JW-transformed wave function:
\begin{align}
|\Psi_\mathrm{JW}\rangle
 &= |-\rangle + \sum_i \eta_i \, a_i^\dagger |-\rangle + \sum_{i<j} \eta_i \, \eta_j \, a_i^\dagger \, a_j^\dagger |-\rangle
\\
 &+ \sum_{i<j<k} \eta_i \, \eta_j \, \eta_k \, a_i^\dagger \, a_j^\dagger \, a_k^\dagger |-\rangle + \ldots
\nonumber
\end{align}

\subsection{Heisenberg Rings}
Now let us turn to simple Heisenberg rings.  Here, we have
\begin{equation}
H = \sum_{\langle pq \rangle} \vec{S}_p \cdot \vec{S}_q
\end{equation}
where only nearest-neighbor pairs $pq$ are coupled, and where we have periodic boundary conditions.  The Hamiltonian favors coupling adjacent sites antiferromagnetically, and when the number of sites is even we obtain a Ne\'el solution.  When the number of sites is odd, however, the system has spin frustration.  This means that in the exact system, the energy per site displays an even-odd alternation.

\begin{table*}[t]
\caption{Energies per site in antiferromagnetically-coupled Heisenberg rings.  We show the exact energy, energy computed with spin mean-field (``Spin MF'') and energies computed with fermionic mean-field on the JW-transformed Hamiltonian in the original basis (``JW-Orig'', with $\theta_k = 0$), the classical basis (``JW-Class'', with $\theta_k = k \, \theta$), and in an optimized basis (``JW-Opt'' in which the angles $\theta_k$ are treated as additional variational parameters.\label{Table:HRing}}
\begin{ruledtabular}
\begin{tabular}{cccccc}
$n$	&	Exact		&	Spin MF	&	JW-Orig	&	JW-Class	&	JW-Opt	\\
\hline
2	&	-0.750 000	&	-0.250 000	&	-0.750 000	&	-0.750 000	&	-0.750 000	\\
3	&	-0.250 000	&	-0.125 000	&	-0.250 000	&	-0.250 000	&	-0.250 000	\\
4	&	-0.500 000	&	-0.250 000	&	-0.478 553	&	-0.478 553	&	-0.478 553	\\
5	&	-0.373 607	&	-0.202 254	&	-0.366 425	&	-0.371 699	&	-0.371 699	\\
6	&	-0.467 129	&	-0.250 000	&	-0.444 444	&	-0.444 444	&	-0.444 444	\\
7	&	-0.407 883	&	-0.225 242	&	-0.398 276	&	-0.402 601	&	-0.402 629	\\
8	&	-0.456 387	&	-0.250 000	&	-0.435 706	&	-0.435 706	&	-0.435 706	\\
9	&	-0.421 922	&	-0.234 923	&	-0.410 695	&	-0.414 627	&	-0.414 654	\\
10	&	-0.451 545	&	-0.250 000	&	-0.433 449	&	-0.433 449	&	-0.433 449	\\
11	&	-0.428 994	&	-0.239 873	&	-0.416 677	&	-0.420 533	&	-0.420 552	\\
12	&	-0.448 949	&	-0.250 000	&	-0.432 726	&	-0.432 726	&	-0.432 726	\\
\end{tabular}
\end{ruledtabular}
\end{table*}

The classical solution is for the spins to be coplanar with equal angles between the adjacent spins.  That is, we write the ground state mean-field as
\begin{equation}
|\chi\rangle = \lvert\downarrow_1 \downarrow_2 \ldots \downarrow_{2n+1}\rangle
\end{equation}
where the direction $\downarrow_k$ on site $k$ could be obtained by rotating the spin in, say, the $xz$ plane by angle $k \, \theta$:
\begin{equation}
\lvert\downarrow_k\rangle = \mathrm{e}^{\mathrm{i} \, k \, \theta \, S_k^y} \lvert\downarrow^{(z)}_k\rangle
\end{equation}
where $\lvert\downarrow^{(z)}_k\rangle$ is the eigenstate of $S_k^z$ with eigenvalue $-1/2$.  For even-membered rings, this angle is simply $\theta = \pi$, i.e. spins on adjacent sites are antialigned.  For odd-membered $n$-site rings, however, the angle between adjacent spins is\cite{Schmidt2003}
\begin{equation}
\theta = \pi \, \left(1 - \frac{1}{n}\right)
\end{equation}
such that the spin vectors on adjacent sites are as near as possible to being antiparallel while having equal angles between them.  That is, we could regard site 1 to be the same as site $n+1$ in an $n$-site system, which means that $n \, \theta$ must be an even multiple of $\pi$.

When we rotate spin $k$ by angle $\theta_k$, the original spins $\left(S_k^x,S_k^y,S_k^z\right)$ are expressed in terms of the rotated spins $\left(\tilde{S}_k^x,\tilde{S}_k^y,\tilde{S}_k^z\right)$ as
\begin{subequations}
\begin{align}
S_k^x &= \cos\left(\theta_k\right) \, \tilde{S}_k^x  - \sin\left(\theta_k\right) \, \tilde{S}_k^z,
\\
S_k^y &= \tilde{S}_k^y,
\\
S_k^z &= \cos\left(\theta_k\right) \, \tilde{S}_k^z +\sin\left(\theta_k\right) \, \tilde{S}_k^x.
\end{align}
\end{subequations}
This means that the Heisenberg Hamiltonian in the $\tilde{S}$ basis is
\begin{align}
H = \sum_{\langle pq\rangle} \Big[&
	\cos(\theta_p - \theta_q) \, \left(\tilde{S}_p^x \, \tilde{S}_q^x + \tilde{S}_p^z \, \tilde{S}_q^z\right)
+	\tilde{S}_p^y \, \tilde{S}_q^y
\\
+&	\sin(\theta_p - \theta_q) \, \left(\tilde{S}_p^x \, \tilde{S}_q^z - \tilde{S}_p^z \, \tilde{S}_q^x\right)
\Big].
\nonumber
\end{align}

Now imagine Jordan-Wigner transforming the Hamiltonian.  We may transform it in the original basis $(\theta_k = 0)$ or in the classical basis $(\theta_k = k \, \theta)$ or even in an optimized basis in which the angles $\theta_k$ are variational parameters.  The fermionic Hamiltonian becomes
\begin{widetext}
\begin{align}
H_\mathrm{JW}
 = \sum_{\langle pq \rangle} \Bigg\{&
		\cos(\theta_p - \theta_q) \, \left[\frac{a_p^\dagger + a_p}{2} \phi_p \, \phi_q \, \frac{a_q^\dagger + a_q}{2} + \bar{n}_p \, \bar{n}_q\right]
	+	\frac{a_p^\dagger - a_p}{2 \, \mathrm{i}} \, \phi_p \, \phi_q \, \frac{a_q^\dagger - a_q}{2 \, \mathrm{i}}
\\
	+&	\sin(\theta_p - \theta_q) \, \left[\frac{a_p^\dagger + a_p}{2} \, \phi_p \, \bar{n}_q 
				- \bar{n}_p \, \frac{a_q^\dagger + a_q}{2} \, \phi_q\right]\Bigg\}.
\nonumber
\end{align}
\end{widetext}
Unless all nearest-neighbor angle pairs $(\theta_p,\theta_q)$ differ by a multiple of $\pi$ so that $\sin(\theta_p - \theta_q) = 0$, the JW-transformed Hamiltonian does not conserve number parity and must be solved with the mean-field methods we have outlined in this work.

\begin{figure}[t]
\includegraphics[width=\columnwidth]{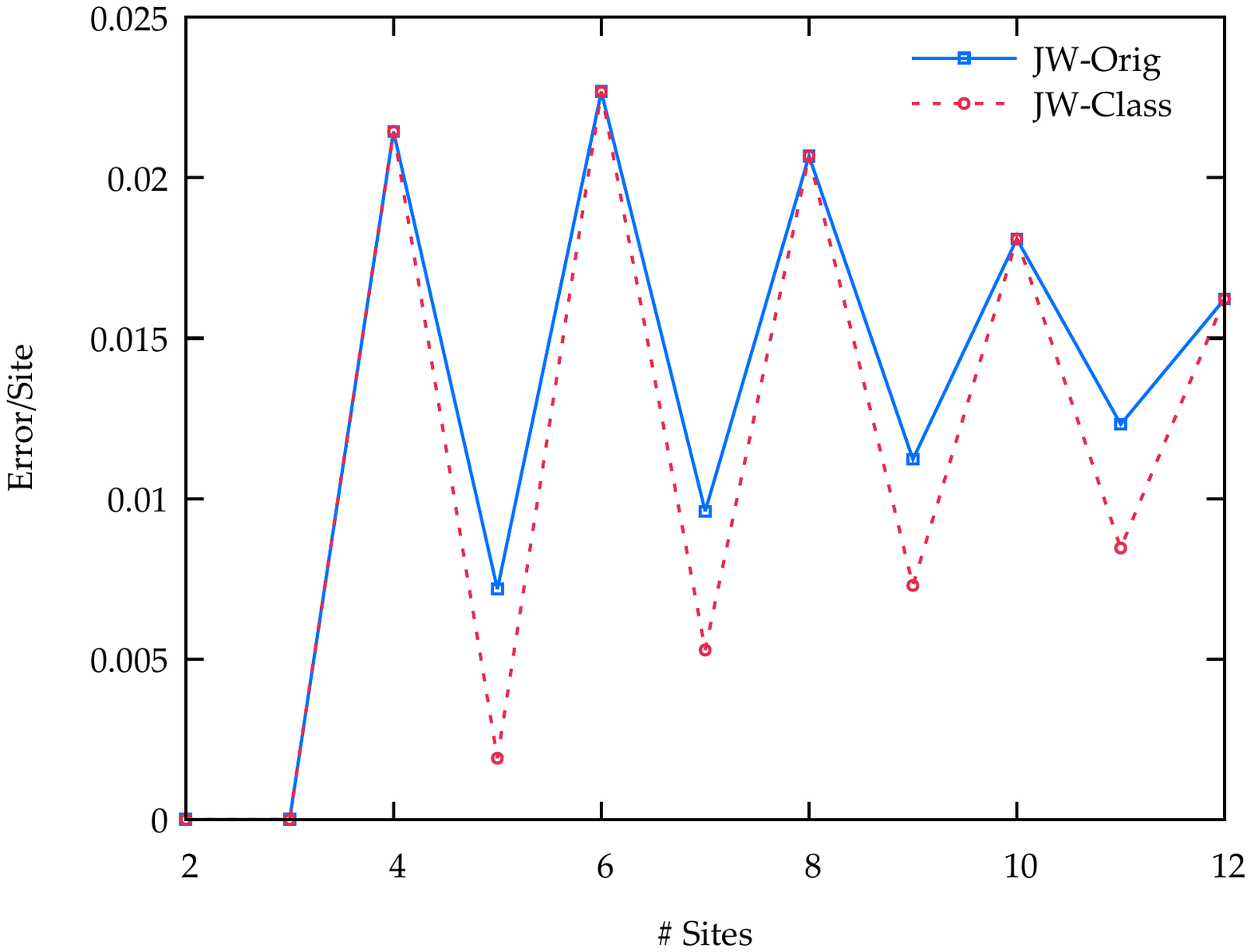}
\caption{Errors per site in $n$-site Heisenberg rings computed in fermionic mean-field after Jordan-Wigner transformation.  We show results in which the transformation is carried out in the original basis (JW-Orig) and in the classical basis (JW-Class).  For odd-numbered rings, the JW calculation in the classical basis is notably more accurate, particularly when the ring is small and the classical and original bases differ strongly.\label{Fig:HeisenbergRing2}}
\end{figure}

Table \ref{Table:HRing} shows the energy per site for $n$-site Heisenberg rings computed exactly, with the spin mean-field and with fermionic mean-field on the JW-transformed Hamiltonian in the original basis, the classical basis, and the optimized basis.  Clearly, we obtain far superior results with the fermionic mean-field methods than we do with a spin mean-field, as we have observed previously \cite{Henderson2022,Henderson2023}.  For non-frustrated systems, it makes no difference whether we do the JW transformation in the original frame, or the classical frame, or even an optimal frame; because the angles $\theta_k$ even when optimized are all $0$ or $\pi$, it suffices in these non-frustrated systems to treat the fermionic Hamiltonian with Hartree-Fock.  For the frustrated rings, however, it is helpful to first solve the problem in a spin mean-field to get the classical solution, and only then do the JW transformation.  Doing so results in a fermionic Hamiltonian which does not have number parity as a symmetry, but we can solve that Hamiltonian at the mean-field level nonetheless, and the error decreases.  Optimizing the basis in which to carry out the JW transformation seems, in this particular case, to not be worth the effort involved.  We note that the improvement from JW-transforming in the classical basis rather than the original basis decreases as the system size increases; this is just because the angles $k \, \theta$ in the classical basis approach $0$ and $\pi$ as we add more sites.  Figure \ref{Fig:HeisenbergRing2} shows error per site in the JW-transformed results, in the original basis and in the classical basis.  We can clearly see the expected even-odd alternation, and the plot readily displays the improvements, for odd-membered rings, obtained by doing the JW transformation in the classical basis.  We have not included results using the optimized basis in this figure as they are visually indistinguishable from those in the classical basis, nor have we included results from spin mean-field simply because the errors are so much larger.

Finally, we note that we have generated the JW results for the frustrated rings in two distinct ways.  On the one hand, we have written the wave function in the form of Eqn. \ref{Eqn:WFN1} and minimized the energy with respect to the parameters $Z_{pq}$ and $t_p$, implemented in a Fock-space full configuration interaction code.  On the other hand, we have followed the procedure outlined in Fukutome's original work on number-parity--violating mean-field solutions \cite{Fukutome1977}.  This process builds a non-interacting effective Hamiltonian which is self-consistently diagonalized, in much the same way as is done for Hartree-Fock or Hartree-Fock-Bogoliubov theory.  Results from the two approaches always agree to the precision in which the calculations were carried out.  In the absence of the Jordan-Wigner strings, HFBF is no more computationally demanding than any other mean-field theory.  The JW strings complicate implementation, and a more efficient direct minimization implementation is underway and will be discussed in due time.

\section{Conclusions}
Mean-field methods have a foundational role in modern computational techniques.  Most familiar fermionic Hamiltonians have particle number symmetry, for which reason a great deal of attention has been devoted to number-conserving mean-field theory (Hartree-Fock) and to correcting its deficiencies by any of a wide variety of methods. The number-violating mean-field theory (Hartree-Fock-Bogoliubov) has seen very little attention in the quantum chemistry literature simply because number symmetry cannot break spontaneously at the mean-field level when the interaction is purely repulsive\cite{Bach1994}, as is the case for the electronic Hamiltonian.  In nuclear physics, the story is somewhat different, and calculations using Hartree-Fock-Bogoliubov (HFB) or number-projected HFB are routine.  Although not as well developed as are post-Hartree-Fock methods, some attention has been paid to correlating HFB with, for example, essentially traditional coupled cluster theory.\cite{Henderson2014,Signoracci2015}

Number parity symmetry, on the other hand, is almost never considered at all.  Indeed, we struggle to think of physical examples for which a number-parity--violating mean-field is essential.  Nor, unfortunately, is there anything to gain from the number-parity projection of our number-parity--violating mean-field wave function, since the end result of doing so is a simple HFB state.  So in some sense, the direct applications for the exotic mean-field theory discussed here seem somewhat limited.

On the other hand, once we transform a spin Hamiltonian into a Hamiltonian of fermions, as we have done here with the Jordan-Wigner transformation, we would not expect that Hamiltonian to have either fermionic number symmetry or fermionic number parity symmetry.  In these cases, the ability to go beyond simple Hartree-Fock or HFB seems quite valuable.  Indeed, as we have seen here, the Jordan-Wigner transformation of a spin system which has, at the spin mean-field level, noncollinear magnetism gives rise to a fermionic Hamiltonian which breaks number parity, and we can take advantage of this fact to obtain superior results using a number-parity--violating mean-field theory.  We have considered only frustrated Heisenberg rings where the spin frustration decreases as the number of sites increases, but these same techniques would be useful in, for example, frustrated spin lattices.  As Moussa notes\cite{Moussa2018}, these ideas may also be useful when even- and odd-number parity sectors of Hilbert space are degenerate.  See, for example, Ref. \onlinecite{Klassen2015}, which considers fermionic number parity breaking by using the JW transformation to transform a fermionic Hamiltonian into a spin Hamiltonian for which the number parity breaking can be treated by spin mean-field methods.  Using the techniques outlined in this work, such fermionic Hamiltonians can be treated directly.

We have only begun to work with these unusual mean-field states.  But because these mean-field states are essentially novel, there is the potential to develop a whole host of new computational techniques based upon these methods.  One could, for example, envision correlating these states with an analog of configuration interaction or coupled cluster theory.  And while the number and number-parity projection of these mean-field states is not of interest, the same need not be true of correlated but parity-violating wave functions.

\section{Data Availability}
The data that support the findings of this study are available within the article.

\begin{acknowledgments}
This work was supported by the U.S. Department of Energy, Office of Basic Energy Sciences, under award DE-FG02-09ER16053.  G.E.S. is a Welch Foundation Chair (C-0036).
\end{acknowledgments}

\appendix
\section{Density Matrices of Hartree-Fock-Bogoliubov-Fukutome States
\label{Sec:DensityMatrices}}
In order to compute with HFBF, we need to calculate its density matrices.  Following Fukutome\cite{Fukutome1977}, we use Wick's theorem except that we may have non-zero expectation values of single fermion operators $a$ and $a^\dagger$.  The guiding notion is that the quasiparticle operators $F$ and $F^\dagger$ obey standard fermionic anticommutation rules, while the annihilation operators $F_p$ annihilate the quasiparticle vacuum $|0_F\rangle$ to the right: $F_p |0_F\rangle = 0$.

Our first step is to invert the transformation from $a$ and $a^\dagger$ to $F$ and $F^\dagger$. Recall that
\begin{subequations}
\begin{align}
F_p^\dagger &= P^\dagger \, \left(U_{ip} \, a_i^\dag + V_{ip} \, a_i + \frac{1}{\sqrt{2}} \, y_p^\star\right)
\\
 &= \left(U_{ip} \, a_i^\dag + V_{ip} \, a_i - \frac{1}{\sqrt{2}} \, y_p^\star\right) \, P,
\nonumber
\\
F_p &= \left(U_{ip}^\star \, a_i + V_{ip}^\star \, a_i^\dagger + \frac{1}{\sqrt{2}} \, y_p\right) \, P
\\
  &= P^\dagger \, \left(U_{ip}^\star \, a_i + V_{ip}^\star \, a_i^\dagger - \frac{1}{\sqrt{2}} \, y_p\right),
\nonumber
\\
P &= z + \sqrt{2} \, x_p^\star \, a_p - \sqrt{2} \, x_p \, a_p^\dagger.
\end{align}
\end{subequations}
We can cast this in matrix form as 
\begin{subequations}
\label{AppB:FukutomeTransformation}
\begin{align}
\begin{pmatrix}	\boldsymbol{F^\dagger}	&	\boldsymbol{F}	&	\frac{1}{\sqrt{2}}	\end{pmatrix} &= P^\dagger \, 
\begin{pmatrix}	\boldsymbol{a^\dagger}	&	\boldsymbol{a}	&	\frac{1}{\sqrt{2}}	\end{pmatrix} \, \boldsymbol{G},
\label{AppB:FukutomeTransformation1}
\\
\begin{pmatrix}	\boldsymbol{F^\dagger}	&	\boldsymbol{F}	&	-\frac{1}{\sqrt{2}}	\end{pmatrix} &= 
\begin{pmatrix}	\boldsymbol{a^\dagger}	&	\boldsymbol{a}	&	-\frac{1}{\sqrt{2}}	\end{pmatrix} \, P \, \boldsymbol{G},
\\
\label{Eqn:DefG}
\boldsymbol{G} &= 
\begin{pmatrix}	\boldsymbol{U}		&	\boldsymbol{V}^\star	&	-\boldsymbol{x}	\\
			\boldsymbol{V}		&	\boldsymbol{U}^\star	&	\boldsymbol{x}^\star	\\
			\boldsymbol{y}^\dagger	&	-\boldsymbol{y}^\mathsf{T}		&	z	
\end{pmatrix}.
\end{align}
\end{subequations}
Note that $\boldsymbol{F^\dagger}$ refers to the row vector of creation operators $F^\dagger$, and not to the hermitian adjoint of the row vector $\boldsymbol{F}$, and similarly for $\boldsymbol{a^\dagger}$.  The equation derived from $1/\sqrt{2}$ in the left-hand-side of Eqn. \ref{AppB:FukutomeTransformation1} is
\begin{subequations}
\begin{align}
\frac{1}{\sqrt{2}}& = P^\dagger \, \left(-x_p \, a_p^\dagger + x_p^\star \, a_p + \frac{1}{\sqrt{2}} \, z\right)
\\
 &= \frac{1}{\sqrt{2}} \, P^\dagger \, \left(z + \gamma\right)
\end{align}
\end{subequations}
and serves to define $P^\dagger$ (by multiplying on the right by $z - \gamma$ and recalling that $z^2 - \gamma^2 = 1$ as a consequence of the unitarity of $\boldsymbol{G}$).

Using unitarity of $\boldsymbol{G}$ and $P$, we can invert the transformation to get
\begin{subequations}
\label{AppB:FukutomeInverseTransformation}
\begin{align}
\begin{pmatrix}	\boldsymbol{a^\dagger}	&	\boldsymbol{a}	&	\frac{1}{\sqrt{2}}	\end{pmatrix} &= P \, 
\begin{pmatrix}	\boldsymbol{F^\dagger}	&	\boldsymbol{F}	&	\frac{1}{\sqrt{2}}	\end{pmatrix}  \, \boldsymbol{G}^\dagger,
\\
\begin{pmatrix}	\boldsymbol{a^\dagger}	&	\boldsymbol{a}	&	-\frac{1}{\sqrt{2}}	\end{pmatrix} &= 
\begin{pmatrix}	\boldsymbol{F^\dagger}	&	\boldsymbol{F}	&	-\frac{1}{\sqrt{2}}	\end{pmatrix}  \,  P^\dagger \,\boldsymbol{G}^\dagger.
\\
\boldsymbol{G}^\dagger &= 
\begin{pmatrix}	\boldsymbol{U}^\dagger	&	\boldsymbol{V}^\dagger	&	\boldsymbol{y}\\
			\boldsymbol{V}^\mathsf{T}	&	\boldsymbol{U}^\mathsf{T}	&	-\boldsymbol{y}^\star	\\
			-\boldsymbol{x}^\dagger	&	\boldsymbol{x}^\mathsf{T}	&	z
\end{pmatrix}.
\end{align}
\end{subequations}
In other words
\begin{subequations}
\begin{align}
a_i^\dagger
	&= P \, \left(U_{ip}^\star \, F_p^\dagger  + V_{ip} \, F_p - \frac{1}{\sqrt{2}} \, x_i^\star\right)
\\
	&= \left(U_{ip}^\star \, F_p^\dagger  + V_{ip} \, F_p + \frac{1}{\sqrt{2}} \, x_i^\star\right) \, P^\dagger,
\nonumber
\\
a_i
	&= P \, \left(V_{ip}^\star \, F_p^\dagger + U_{ip} \, F_p + \frac{1}{\sqrt{2}} \, x_i\right)
\\
	&= \left(V_{ip}^\star \, F_p^\dagger + U_{ip} \, F_p - \frac{1}{\sqrt{2}} \, x_i\right) \, P^\dagger,
 \nonumber
 \\
 \frac{1}{\sqrt{2}}
 	&=	P \, \left(y_p \, F_p^\dagger - y_p^\star \, F_p + \frac{1}{\sqrt{2}} \, z\right).
 \end{align}
\end{subequations}
Unitarity of $P$ means that the last equation gives us
\begin{equation}
P^\dagger = z + \sqrt{2} \, y_p \, F_p^\dagger - \sqrt{2} \, y_p^\star \, F_p.
\end{equation}
We can thus write $a$ and $a^\dagger$ in terms of $F$ and $F^\dagger$.

Our strategy now is to evaluate one-body density matrices with $P^\dagger$ to the right of the first operator and $P$ to the left of the second.  For example, to evaluate $\langle a_i^\dagger \, a_j\rangle$, we write
\begin{subequations}
\begin{align}
a_i^\dagger &= \left(U_{ip}^\star \, F_p^\dagger  + V_{ip} \, F_p + \frac{1}{\sqrt{2}} \, x_i^\star\right) \, P^\dagger,
\\
a_j &=  P \, \left(V_{jq}^\star \, F_q^\dagger + U_{jq} \, F_q + \frac{1}{\sqrt{2}} \, x_j\right).
\end{align}
\end{subequations}
Using that $P^\dagger \, P = 1$ and that $F_p^\dagger$ annihilates the vacuum to the left and $F_q$ to the right, we find
\begin{subequations}
\begin{align}
\langle a_i^\dagger \, a_j\rangle 
 &=\langle (V_{ip} \, F_p + \frac{1}{\sqrt{2}} \, x_i^\star) \, (V_{jq}^\star \, F_q^\dagger + \frac{1}{\sqrt{2}} \, x_j)\rangle
\\
 &= V_{ip} \, V_{jp}^\star + \frac{1}{2} \, x_i^\star \, x_j.
\end{align}
\end{subequations}

Similarly, for $\langle a_i^\dagger \, a_j^\dagger\rangle$, we use
\begin{equation}
a_j^\dagger = P \, \left(U_{jq}^\star \, F_q^\dagger  + V_{jq} \, F_q - \frac{1}{\sqrt{2}} \, x_q^\star\right)
\end{equation}
from which we extract
\begin{subequations}
\begin{align}
\langle a_i^\dagger \, a_j^\dagger\rangle
	&=	\langle (V_{ip} \, F_p + \frac{1}{\sqrt{2}} \, x_i^\star) \, (U_{jq}^\star \, F_q^\dagger - \frac{1}{\sqrt{2}} \, x_j^\star) \rangle
\\
	&=	V_{ip} \, U_{jp}^\star - \frac{1}{2} \, x_i^\star \, x_j^\star.
\end{align}
\end{subequations}
Of course 
\begin{equation}
\langle a_j \, a_i \rangle = \langle a_i^\dagger \, a_j^\dagger\rangle^\star.
\end{equation}

For the single-fermion operators, we have to account for $P$ or $P^\dagger$.  Noting that
\begin{subequations}
\begin{align}
P^\dagger |0_F\rangle
	&=	\left(z + \sqrt{2} \, y_q \, F_q^\dagger - \sqrt{2} \, y_q^\star \, F_q\right) |0_F\rangle
\\
	&=	\left(z + \sqrt{2} \, y_q \, F_q^\dagger\right) |0_F\rangle,	
\end{align}
\end{subequations}
we find
\begin{subequations}
\begin{align}
\langle a_i^\dagger\rangle
	&=	\langle (V_{ip} \, F_p + \frac{1}{\sqrt{2}} \, x_i^\star) \, (z + \sqrt{2} \, y_q \, F_q^\dagger) \rangle
\\
	&=	\frac{1}{\sqrt{2}} \, z \, x_i^\star + \sqrt{2} \, V_{ip} \, y_p.
\end{align}
\end{subequations}

Note that we can simplify the density matrices by using the unitarity of $\boldsymbol{G}$.  By writing $\boldsymbol{G} \, \boldsymbol{G}^\dagger = \boldsymbol{1}$, we find that
\begin{subequations}
\label{Eqn:GGStar}
\begin{align}
x_i^\star \, x_j  &= \delta_{ij} - U_{ip}^\star  \, U_{jp}-  V_{ip} \, V_{jp}^\star,
\\
x_i^\star \, x_j^\star &=  V_{ip} \, U_{jp}^\star  + V_{jp} \, U_{ip}^\star,
\\
z \, x_i^\star &= U_{ip}^\star \, y_p^\star - V_{ip} \, y_p.
\end{align}
\end{subequations}
This allows us to write
\begin{subequations}
\begin{align}
\langle a_i^\dagger \, a_j\rangle &=\frac{1}{2} \, \delta_{ij} + \frac{1}{2} \, V_{ip} \, V_{jp}^\star - \frac{1}{2} \, U_{ip}^\star \, U_{jp},
\\
\langle a_i^\dagger \, a_j^\dagger \rangle &= \frac{1}{2} \, \left(V_{ip} \, U_{jp}^\star - V_{jp} \, U_{ip}^\star\right),
\\
\langle a_i^\dagger \rangle &= \frac{1}{\sqrt{2}} \, \left(U_{ip}^\star \, y_p^\star + V_{ip} \, y_p\right).
\end{align}
\end{subequations}

\section{Relation Between the FukutomeTransformation and a Simplified Wave Function
\label{Sec:FukutomeProof}}
In this appendix, we prove that quasiparticle operators $F$
associated with the unitary transformation matrix $\boldsymbol{G}$ of Eqn. \ref{Eqn:DefG}
with $\det(\boldsymbol{G}) = 1$ annihilate the wave function
\begin{subequations}
\begin{align}
|\Phi(Z,t)\rangle &= \left[1 + t_p \, \alpha_p^\dagger(Z)\right] \, |\Phi(Z)\rangle,
\label{Def:PhiZT}
\\
|\Phi(Z)\rangle &= \mathrm{e}^Z |-\rangle,
\\
Z &= \sum_{p>q} Z_{pq} \, a_p^\dagger \, a_q^\dagger,
\end{align}
\end{subequations}
where $|-\rangle$ is the physical vacuum and where $\alpha_p(Z)$ annihilates $|\Phi(Z)\rangle$.  This means that $|\Phi(Z,t)\rangle$ is the vacuum for Fukutome's $SO(2M+1)$ representation.  We also show how we can extract the Fukutome transformation matrix $\boldsymbol{G}$ given $\boldsymbol{t}$ and the Bogoliubov transformation associated with the quasiparticle operators $\alpha(Z)$ and $\alpha^\dagger(Z)$, and vice versa.

For brevity, we will suppress the $Z$-dependence of the quasiparticle operators $\alpha(Z)$ and $\alpha^\dagger(Z)$.  They are given in terms of the bare fermion operators $a$ and $a^\dagger$ via
\begin{subequations}
\begin{align}
\begin{pmatrix} \boldsymbol{\alpha^\dagger} & \boldsymbol{\alpha}\end{pmatrix}
 &= \begin{pmatrix} \boldsymbol{a^\dagger} & \boldsymbol{a}\end{pmatrix} \, \boldsymbol{W}_0,
\\
\boldsymbol{W}_0 &= \begin{pmatrix}    \boldsymbol{U}_0 & \boldsymbol{V}_0^\star\\    \boldsymbol{V}_0 & \boldsymbol{U}_0^\star  \end{pmatrix},
\end{align}
\end{subequations}
where $\boldsymbol{W}_0$ is unitary with determinant 1; we have aleady seen that $\boldsymbol{W}_0$ must be unitary, and its determinant can be proven from the Bloch-Messiah decomposition of $\boldsymbol{W}_0$.\cite{Bloch1962}  We define
\begin{equation}
\boldsymbol{G}_0 = \begin{pmatrix} \boldsymbol{W}_0 & \boldsymbol{0}\\    \boldsymbol{0} & 1  \end{pmatrix}
\end{equation}
which of course is also unitary with determinant 1.  Note that $\boldsymbol{W}_0$ is related to the antisymmetric matrix $\boldsymbol{Z}$ defining the Thouless transformation $Z$ via $\boldsymbol{Z} = \left(\boldsymbol{V}_0\, \boldsymbol{U}_0^{-1}\right)^\star$. We can show that
\begin{subequations}
\begin{align}
  \boldsymbol{U}_0 &= \boldsymbol{L}^{-\dag}\, \boldsymbol{C},\\
  \boldsymbol{V}_0 &= \boldsymbol{Z}^\star\, \boldsymbol{L}^{-\dag}\, \boldsymbol{C},
\end{align}
\end{subequations}
where $\boldsymbol{L}$ is defined by the Cholesky decomposition
\begin{equation}
  \boldsymbol{L}\, \boldsymbol{L}^\dag
  = \boldsymbol{1} + \boldsymbol{Z}\, \boldsymbol{Z}^\dag
\end{equation}
and $\boldsymbol{C}$ is an arbitrary unitary matrix.

\begin{figure}[t!]
  \centering
  \begin{tikzcd}[column sep=4.5em, row sep=4.5em]
    (a^\dag, a) \arrow[rd, "{\boldsymbol{G}}"] \arrow{r}{\boldsymbol{G}_0} & (\alpha^\dag, \alpha) \arrow{d}{\tilde{\boldsymbol{G}}}\\
                                                                           & \left(F^\dag, F\right)
  \end{tikzcd}
  \caption{Transformations between different sets of fermionic operators.}
  \label{fig:transformations}
\end{figure}

Our first task is to establish that Fukutome quasiparticle operators $F_p$ annihilate $|\Phi(Z,t)\rangle$ while satisfying the constraints that $\boldsymbol{G}$ is unitary with determinant 1.   It simplifies the algebra to write $F_p$ in terms of the quasiparticle operators $\alpha$ and $\alpha^\dagger$ instead of the bare fermion operators $a$ and $a^\dagger$.  We have
\begin{subequations}
\begin{align}
F_p &= \tilde{P}^\dagger \left(\tilde{U}_{qp}^\star \, \alpha_q + \tilde{V}_{qp}^\star \, \alpha_q^\dagger - \frac{1}{\sqrt{2}} \, y_p\right),
\label{Eqn:FukutomeAnnihilationOperator}
\\
\tilde{P}^\dagger &= z - \sqrt{2} \, \left(\tilde{x}_p^\star \, \alpha_p - \tilde{x}_p \, \alpha_p^\dagger\right),
\end{align}
\end{subequations}
where the coefficients $\tilde{\boldsymbol{U}}$, $\tilde{\boldsymbol{V}}$, $\tilde{\boldsymbol{x}}$, and $\boldsymbol{y}$ are associated with the transformation matrix
\begin{equation}
\tilde{\boldsymbol{G}}
  = \begin{pmatrix}
    \tilde{\boldsymbol{U}} & \tilde{\boldsymbol{V}}^\star & -\tilde{\boldsymbol{x}}\\
    \tilde{\boldsymbol{V}} & \tilde{\boldsymbol{U}}^\star & \tilde{\boldsymbol{x}}^\star\\
    \boldsymbol{y}^\dag & -\boldsymbol{y}^\T & z
  \end{pmatrix}.
 \label{Def:GTilde}
\end{equation}
We recover $\boldsymbol{G}$ from $\boldsymbol{G}_0$ and $\tilde{\boldsymbol{G}}$ as
\begin{equation}
\boldsymbol{G} = \boldsymbol{G}_0 \, \tilde{\boldsymbol{G}}.
\end{equation}
Figure \ref{fig:transformations} summarizes the relations among $\boldsymbol{G}_0$, $\tilde{\boldsymbol{G}}$, and $\boldsymbol{G}$.

Now, acting $F_p$ on $|\Phi(Z,t)\rangle$, we obtain
\begin{widetext}
\begin{equation}
  F_p \, |\Phi(Z,t)\rangle
  = \tilde{P}^\dagger\, \left[
		\left(\tilde{U}_{qp}^\star\, t_q - \frac{1}{\sqrt{2}}\, y_p\right)
	+	\left(\tilde{V}_{qp}^\star - \frac{1}{\sqrt{2}}\, y_p t_q\right) \, \alpha_q^\dag
	+	\tilde{V}_{qp}^\star\, t_r\, \alpha_q^\dag\, \alpha_r^\dag
	\right] \, |\Phi(Z)\rangle.
\end{equation}
\end{widetext}
The first two terms on the right-hand-side vanish when
\begin{subequations}
\label{Eqn:yVConstraints}
\begin{align}
  \label{Eqn:Defy}
  \boldsymbol{y}
  &= \sqrt{2}\, \tilde{\boldsymbol{U}}^\dag\, \boldsymbol{t},
  \\
  \label{Eqn:DefVt}
  \tilde{\boldsymbol{V}}
  &= \frac{1}{\sqrt{2}}\, \boldsymbol{t}^\star\, \boldsymbol{y}^\dag
  = \boldsymbol{t}^\star\, \boldsymbol{t}^\dag\, \tilde{\boldsymbol{U}}.
\end{align}
\end{subequations}
The last term, with $\alpha_q^\dag\, \alpha_r^\dag$, also drops out, because
\begin{equation}
  \tilde{V}_{qp}^\star\, t_r\, \alpha_q^\dag\, \alpha_r^\dag
  = \frac{1}{\sqrt{2}}\, y_p\, t_q\, t_r\, \alpha_q^\dag\, \alpha_r^\dag
  = 0.
\end{equation}
That the left-hand-side vanishes is because $t_q \, t_r \, \alpha_q^\dagger \, \alpha_r^\dagger = t^\dagger \, t^\dagger$ where $t^\dagger = t_q \, \alpha_q^\dagger$ is a fermionic creation operator.

This much serves to show that we can find operators $F_p$ which annihilate $|\Phi(Z,t)\rangle$.  We have also determined $\boldsymbol{y}$ and $\tilde{\boldsymbol{V}}$ in terms of $\tilde{\boldsymbol{U}}$ and $\boldsymbol{t}$.  It remains to find $\tilde{\boldsymbol{U}}$, $\tilde{\boldsymbol{x}}$, and $z$ such that $\tilde{\boldsymbol{G}}$  satisfies the constraints of Fukutome's $SO(2M+1)$ representation, i.e. $\tilde{\boldsymbol{G}}$ is unitary with determinant 1.

Unitarity entails
\begin{subequations}
\begin{align}
  \label{Eqn:ConstrainA}
  \tilde{\boldsymbol{U}}^\dag \, \tilde{\boldsymbol{U}} + \tilde{\boldsymbol{V}}^\dag \, \tilde{\boldsymbol{V}} + \boldsymbol{y} \, \boldsymbol{y}^\dag
  &= \mathbf{1},
  \\
  \label{Eqn:ConstrainB}
  \tilde{\boldsymbol{U}}^\T \, \tilde{\boldsymbol{V}} + \tilde{\boldsymbol{V}}^\T \, \tilde{\boldsymbol{U}}  - \boldsymbol{y}^\star \, \boldsymbol{y}^\dag
  &= \mathbf{0},
  \\
  \label{Eqn:ConstrainC}
  \tilde{\boldsymbol{U}}^\dag \, \tilde{\boldsymbol{x}} - \tilde{\boldsymbol{V}}^\dag \, \tilde{\boldsymbol{x}}^\star - z \, \boldsymbol{y}
  &= \boldsymbol{0},
  \\
  \label{Eqn:ConstrainD}
  2\, \tilde{\boldsymbol{x}}^\dagger \, \tilde{\boldsymbol{x}} + z^2
  &= 1.
\end{align}
\end{subequations}
With the aid of Eqn. \ref{Eqn:yVConstraints} we can verify that Eqn. \ref{Eqn:ConstrainB} is already satisfied, while Eqns. \ref{Eqn:ConstrainA} and \ref{Eqn:ConstrainC} reduce to
\begin{subequations}
\begin{align}
  \label{Eqn:ConstrainAA}
  \tilde{\boldsymbol{U}}^\dag
  \left[
    \boldsymbol{1} + \left(2 + \boldsymbol{t}^\dag\, \boldsymbol{t}\right)\,
    \boldsymbol{t}\, \boldsymbol{t}^\dag
  \right] \, \tilde{\boldsymbol{U}}
  &= \boldsymbol{1},
  \\
  \label{Eqn:ConstrainCC}
  \tilde{\boldsymbol{U}}^\dag \left(
    \tilde{\boldsymbol{x}}
    - \boldsymbol{t}\, \boldsymbol{t}^\T\, \tilde{\boldsymbol{x}}^\star
    - \sqrt{2}\, z\, \boldsymbol{t}
  \right)
  &= \boldsymbol{0}.
\end{align}
\end{subequations}

Since $\boldsymbol{1} + \left(2 + \boldsymbol{t}^\dag\, \boldsymbol{t}\right)\, \boldsymbol{t}\, \boldsymbol{t}^\dag$ is a rank-1 update of the identity, we can readily see that
\begin{equation}
\label{Eqn:DefL}
\boldsymbol{1} + \left(2 + \boldsymbol{t}^\dag\, \boldsymbol{t}\right)\, \boldsymbol{t}\, \boldsymbol{t}^\dag = \left(\boldsymbol{1} + \boldsymbol{t}\, \boldsymbol{t}^\dag\right)^2,
\end{equation}
and Eqn. \ref{Eqn:ConstrainAA} implies
\begin{equation}
\label{Eqn:DefUt}
\tilde{\boldsymbol{U}} = \left(\boldsymbol{1} + \boldsymbol{t}\, \boldsymbol{t}^\dag\right)^{-1} \, \boldsymbol{S} = \left[\boldsymbol{1} - \left(1 + \boldsymbol{t}^\dag\, \boldsymbol{t}\right)^{-1}\, \boldsymbol{t}\, \boldsymbol{t}^\dag\right] \, \boldsymbol{S},
\end{equation}
where $\boldsymbol{S}$ is an arbitrary unitary matrix.

Because $\tilde{\boldsymbol{U}} ^\dagger$ is invertible, Eqn. \ref{Eqn:ConstrainCC} yields
\begin{equation}
\tilde{\boldsymbol{x}} = \left(\boldsymbol{t}^\T\, \tilde{\boldsymbol{x}}^\star + \sqrt{2} z\right)\, \boldsymbol{t}.
\end{equation}
We see that
\begin{equation}
\label{Eqn:Defw}
\tilde{\boldsymbol{x}} = w \, \boldsymbol{t}
\end{equation}
for some complex number $w$.  We insert this relation into Eqns. \ref{Eqn:ConstrainD} and \ref{Eqn:ConstrainCC} and solve for $z$ and $w$, which gives
\begin{subequations}
\label{Eqn:Defz1w1}
\begin{align}
  \label{Eqn:Defz1}
  z &= \pm \frac{1 - \boldsymbol{t}^\dag\, \boldsymbol{t}}{1 + \boldsymbol{t}^\dag\, \boldsymbol{t}},
  \\
  w &= \pm \frac{\sqrt{2}}{1 + \boldsymbol{t}^\dag\, \boldsymbol{t}},
\end{align}
\end{subequations}
and hence
\begin{equation}
\tilde{\boldsymbol{x}}  = \pm \frac{\sqrt{2}}{1 + \boldsymbol{t}^\dag\, \boldsymbol{t}}\, \boldsymbol{t}.
\end{equation}
Note that we have used $\boldsymbol{t} \neq \boldsymbol{0}$; otherwise, $\tilde{\boldsymbol{G}}$ reduces to a standard Bogoliubov transformation with $\tilde{\boldsymbol{x}} = \boldsymbol{0} = \boldsymbol{y}$ and $z = 1$.

At this point, we have shown that $\tilde{\boldsymbol{G}}$ is unitary.  We must still check its determinant.  To do so, we temporarily assume $z \ne 0$, leading to
\begin{align}
  \nonumber
  \det(\tilde{\boldsymbol{G}})
  &= z\, \det\left(
    \begin{pmatrix}
      \tilde{\boldsymbol{U}} & \tilde{\boldsymbol{V}}^\star\\
      \tilde{\boldsymbol{V}} & \tilde{\boldsymbol{U}}^\star
    \end{pmatrix}
    - \begin{pmatrix}
      -\tilde{\boldsymbol{x}} \\ \tilde{\boldsymbol{x}}^\star
    \end{pmatrix}
    z^{-1}
    \begin{pmatrix}
      \boldsymbol{y}^\dag & -\boldsymbol{y}^\T
    \end{pmatrix}
  \right)\\
  \nonumber
  &= z\, \det\left(
    \begin{pmatrix}
      \boldsymbol{X} & \boldsymbol{Y}^\star\\
      \boldsymbol{Y} & \boldsymbol{X}^\star
    \end{pmatrix}
    \,
    \begin{pmatrix}
      \boldsymbol{S} & \boldsymbol{0}\\
      \boldsymbol{0} & \boldsymbol{S}^\star
    \end{pmatrix}
  \right)\\
  \nonumber
  &= z \, \det(\boldsymbol{M})\, |\det(\boldsymbol{S})|^2\\
  &= z \, \det(\boldsymbol{M}),
  \label{Eqn:DetGt}
\end{align}
where
\begin{equation}
  \boldsymbol{M} =
  \begin{pmatrix}
    \boldsymbol{X} & \boldsymbol{Y}^\star\\
    \boldsymbol{Y} & \boldsymbol{X}^\star
  \end{pmatrix},
\end{equation}
and
\begin{subequations}
\begin{align}
  \nonumber
  \boldsymbol{X}
  &= \left(
    \tilde{\boldsymbol{U}} + z^{-1}\, \tilde{\boldsymbol{x}}\, \boldsymbol{y}^\dag
  \right)\, \boldsymbol{S}^\dag\\
  &= \boldsymbol{1} + (1 - \boldsymbol{t}^\dag\, \boldsymbol{t})^{-1}\, \boldsymbol{t}\, \boldsymbol{t}^\dag,\\
  \nonumber
  \boldsymbol{Y}
  &= \left(
    \tilde{\boldsymbol{V}} - z^{-1}\, \tilde{\boldsymbol{x}}^\star\, \boldsymbol{y}^\dag
  \right)\, \boldsymbol{S}^\T\\
  &= - (1 - \boldsymbol{t}^\dag\, \boldsymbol{t})^{-1}\, \boldsymbol{t}^\star\, \boldsymbol{t}^\dag.
\end{align}
\end{subequations}

Observationally, $\boldsymbol{M}$ has a nondegenerate eigenvalues $\frac{1 + \boldsymbol{t}^\dag\, \boldsymbol{t}}{1 - \boldsymbol{t}^\dag\, \boldsymbol{t}}$ with the associated eigenvector $\boldsymbol{v} = \frac{1}{\sqrt{2 \boldsymbol{t}^\dag \boldsymbol{t}}}\, \begin{pmatrix}   \boldsymbol{t}\\ -\boldsymbol{t}^* \end{pmatrix}$, while the remaining $2M-1$ eigenvalues are degenerate and are equal to 1. To show this analytically, we define
\begin{align}
  \boldsymbol{R}
  &= \boldsymbol{M}
  - \frac{1 + \boldsymbol{t}^\dag\, \boldsymbol{t}}{1 - \boldsymbol{t}^\dag\, \boldsymbol{t}}\, \boldsymbol{v}\, \boldsymbol{v}^\dag\\
  &= \begin{pmatrix}
    \boldsymbol{1} - (2\, \boldsymbol{t}^\dag\, \boldsymbol{t})^{-1}\, \boldsymbol{t}\, \boldsymbol{t}^\dag &
    (2\, \boldsymbol{t}^\dag\, \boldsymbol{t})^{-1}\, \boldsymbol{t}\, \boldsymbol{t}^\T \\
    (2\, \boldsymbol{t}^\dag\, \boldsymbol{t})^{-1}\, \boldsymbol{t}^\star\, \boldsymbol{t}^\dag &
    \boldsymbol{1} - (2\, \boldsymbol{t}^\dag\, \boldsymbol{t})^{-1}\, \boldsymbol{t}^\star\, \boldsymbol{t}^\T &
  \end{pmatrix}.
\end{align}
It is readily seen that $\boldsymbol{R}$ is idempotent and
\begin{equation}
  \text{tr}(\boldsymbol{R}) = 2M - 1.
\end{equation}
This implies that the orthogonal complement of $\boldsymbol{v}$ is a $(2M-1)$-dimensional eigenspace of $\boldsymbol{M}$ associated with the eigenvalue 1. We therefore have
\begin{align}
  \det(\boldsymbol{M})
  = \frac{1 + \boldsymbol{t}^\dag\, \boldsymbol{t}}{1 - \boldsymbol{t}^\dag\, \boldsymbol{t}}
  \cdot 1^{2M-1}
  = \frac{1 + \boldsymbol{t}^\dag\, \boldsymbol{t}}{1 - \boldsymbol{t}^\dag\, \boldsymbol{t}}.
  \label{Eqn:DetM}
\end{align}
Combining Eqns. \ref{Eqn:DetGt} and \ref{Eqn:DetM} and inserting Eqn. \ref{Eqn:Defz1},  we see that
\begin{equation}
\det(\tilde{\boldsymbol{G}}) = z \, \frac{1 + \boldsymbol{t}^\dagger \, \boldsymbol{t}}{1 - \boldsymbol{t}^\dagger \, \boldsymbol{t}} = \pm 1,
\end{equation}
which has two implications. First, since this expression is independent of $z$, it should also hold in the $z \to 0$ limit.  Second, to ensure $\det(\tilde{\boldsymbol{G}}) = 1$, we need to choose the positive sign in Eqn. \ref{Eqn:Defz1w1}, i.e.,
\begin{subequations}
\label{Eqn:Defzxt}
\begin{align}
z &= \frac{1 - \boldsymbol{t}^\dag\, \boldsymbol{t}}{1 + \boldsymbol{t}^\dag\, \boldsymbol{t}},
\\
\label{Eqn:Defxt}
\tilde{\boldsymbol{x}} &=  \frac{\sqrt{2}}{1 + \boldsymbol{t}^\dag\, \boldsymbol{t}}\, \boldsymbol{t}.
\end{align}
\end{subequations}

Together with
\begin{subequations}
\label{Eqn:DefUtVty}
\begin{align}
  \tilde{\boldsymbol{U}}
  &= \left[\boldsymbol{1} - \left(1 + \boldsymbol{t}^\dag\, \boldsymbol{t}\right)^{-1}\, \boldsymbol{t}\, \boldsymbol{t}^\dag\right] \, \boldsymbol{S},
  \\
  \tilde{\boldsymbol{V}}
  &= \left(1 + \boldsymbol{t}^\dag\, \boldsymbol{t}\right)^{-1}\, \boldsymbol{t}^\star\, \boldsymbol{t}^\dag \, \boldsymbol{S},
  \\
  \boldsymbol{y}
  &= \sqrt{2}\, \left(1 + \boldsymbol{t}^\dag\, \boldsymbol{t}\right)^{-1}\, \boldsymbol{S}^\dag \, \boldsymbol{t},
\end{align}
\end{subequations}
these results establish that the state $|\Phi(Z,t)\rangle$ is annihilated by a Fukutome-style annihilation operator $F_p$ for which $\tilde{\boldsymbol{G}}$ (and hence $\boldsymbol{G}$) is unitary and has determinant 1; ergo, $|\Phi(Z,t)\rangle$ is the vacuum for Fukutome's $SO(2M+1)$ construction.

We have seen that for a given $\boldsymbol{t}$, $\tilde{\boldsymbol{G}}$ is of the form
\begin{align}
  \tilde{\boldsymbol{G}} = 
  \begin{pmatrix}
    \boldsymbol{1} - \xi\, \boldsymbol{t}\, \boldsymbol{t}^\dag &
    \xi\, \boldsymbol{t}\, \boldsymbol{t}^\T &
    -\sqrt{2}\, \xi\, \boldsymbol{t} \\
    \xi\, \boldsymbol{t}^\T\, \boldsymbol{t}^\dag &
    \boldsymbol{1} - \xi\, \boldsymbol{t}^\star\, \boldsymbol{t}^\T &
    \sqrt{2}\, \xi\, \boldsymbol{t}^\star \\
    \sqrt{2}\, \xi\, \boldsymbol{t}^\dag &
    -\sqrt{2}\, \xi\, \boldsymbol{t}^\T &
    \xi\, (1 - \boldsymbol{t}^\dag\, \boldsymbol{t})
  \end{pmatrix}\,
  \begin{pmatrix}
    \boldsymbol{S} &                      & \\
                   & \boldsymbol{S}^\star & \\
                   &                      & 1
  \end{pmatrix},
\label{Eqn:FinalGTilde}
\end{align}
where $\xi = (1 + \boldsymbol{t}^\dag\, \boldsymbol{t})^{-1}$ and the unitary matrix $\boldsymbol{S}$ is physically inconsequential since it only mixes fermionic operators $F^\dag$ among themselves.  We can thus construct $\boldsymbol{G}$ given $\boldsymbol{t}$ and $\boldsymbol{W}_0$, recalling that $\boldsymbol{G} = \boldsymbol{G}_0 \, \tilde{\boldsymbol{G}}$.

We can also do the converse: Given $\boldsymbol{G}$, we can determine $\boldsymbol{t}$ and, up to an overall unitary transformation, the Bogoliubov coefficients $\boldsymbol{W}_0$.  To do so, we start by setting
$\boldsymbol{S} = \boldsymbol{1}$ and
\begin{equation}
  \boldsymbol{t} = \frac{\sqrt{2}}{1 + z}\, \boldsymbol{y}.
\end{equation}
We can then build $\tilde{\boldsymbol{x}}$, $\tilde{\boldsymbol{U}}$, and $\tilde{\boldsymbol{V}}$ from Eqns. \ref{Eqn:Defzxt} and \ref{Eqn:DefUtVty}, resulting in
\begin{subequations}
\begin{align}
\tilde{\boldsymbol{U}} &= \boldsymbol{1} - \frac{1}{1 + z}\, \boldsymbol{y}\, \boldsymbol{y}^\dag,
\\
\tilde{\boldsymbol{V}} &= \frac{1}{1 + z}\, \boldsymbol{y}^\star\, \boldsymbol{y}^\dag,
\\
\tilde{\boldsymbol{x}} &= \boldsymbol{y}.
\end{align}
\end{subequations}
It can be verified that $\boldsymbol{t}$ and $\tilde{\boldsymbol{G}}$ defined in this way are consistent.  We can then obtain $\boldsymbol{G}_0$, and therefore $\boldsymbol{W}_0$, by
\begin{equation}
\boldsymbol{G}_0 = \boldsymbol{G}\, \tilde{\boldsymbol{G}}^\dag.
\end{equation}
The overall unitary transformation aluded to in defining $\boldsymbol{G}_0$ arises because $\tilde{\boldsymbol{G}}$ is defined only up to such a transformation (cf. Eqn. \ref{Eqn:FinalGTilde}).

We conclude this appendix by pointing out that the number of independent real parameters in $\boldsymbol{t}$ and $\boldsymbol{W}_0$ are $2M$ and $M\,(2M-1)$, respectively.  The latter can be determined by the fact that Bogoliubov transformations form the $SO(2M)$ group. Therefore, the total number of independent real parameters in our parametrization of a number-parity-breaking mean-field is the same as that of Fukutome's, as it should be.

\bibliography{JordanWignerBib}

\end{document}